# Computing Light Transport Gradients using the Adjoint Method

Jos Stam, Graphics Researcher, NVIDIA.

## Abstract

This paper proposes a new equation from continuous adjoint theory to compute the gradient of quantities governed by the Transport Theory of light. Unlike discrete gradients ala *autograd*[1], which work at the code level, we first formulate the continuous theory and then discretize it. The key insight of this paper is that computing gradients in Transport Theory is akin to computing the importance, a quantity adjoint to radiance that satisfies an adjoint equation. Importance tells us where to look for light that matters. This is one of the key insights of this paper. In fact, this mathematical journey started from a whimsical thought that these adjoints might be related. Computing gradients is therefore no more complicated than computing the importance field. This insight and the following paper hopefully will shed some light on this complicated problem and ease the implementations of gradient computations in existing path tracers.

## 1. Introduction

In this paper we present a general framework for computing gradients in the context of light propagation. Gradients are of central importance in the fields of machine learning, computer vision and computer graphics. Often, we need to invert a simulation like a rendering to recover *hidden* control parameters. For smooth problems the gradient is a key instrument in methods such as gradient descent or quasi-Newton iteration. This paper is concerned with computing the gradient of a solution to a transport equation. In order to achieve this goal, we derive a continuous adjoint equation for the gradient of the radiance. This equation is a generalization of the usual **backpropagation** algorithm popular in deep learning. We show that the adjoint equation for the gradient is almost identical to the adjoint equation for the *importance* in transport theory. The only difference is a different source term that is equal to the initial gradient of the cost function with respect to the radiance field.

The method of computation is akin to a bi-directional Monte Carlo solution using radiance and importance. First the transport equation is solved for the radiance *forward* from the light sources to the receivers (camera/eye). Then the adjoint transport equation is solved **backwards** from the receiver for the adjoint of the gradient of radiance similarly to the importance. As the propagation progresses backwards, we update the gradients of the cost function with respect to the controls acting at that point in the path. The reader familiar with backpropagation in deep learning will appreciate the analogy with forward and backward propagation in neural networks. Keep this in mind when reading this paper. In fact, the adjoint theory of optimization is the backbone of backpropagation.

---

[1] Autograd is just one of many packages out there that computes differentials at the code level. See https://pytorch.org/tutorials/beginner/blitz/autograd_tutorial.html for more details and [2] for an excellent introduction to Automatic Differentiation.

We contrast our approach with a purely *autograd* style of computing the gradient. Indeed, one could automatically translate the code of a renderer into its adjoint (or reverse) version and thus obtain the gradient. This approach is known as **D.T.O: *Discretize Then Optimize***. On the other hand, our approach falls in the category of **O.T.D: *Optimize Then Discretize*** methods. We derive the adjoint equation in the continuous setting and then discretize and reuse a standard implementation of a path tracer renderer. This is because the adjoint equation is almost identical to the computation of importance. Of course, the differentials appearing in the transport process that depend on the controls must be differentiated, either analytically or using automatic differentiation.

This paper does not address the problem of smoothing non-continuous terms in the transport equation. The problem of handling discontinuities is orthogonal to the approach taken in this paper. We think that uncovering the mathematical structure in a smooth setting sheds another light on the problem and might lead to simplifications, insights and better implementations. We assume in our derivations that each function is differentiable. For non-differentiable terms some regularization or some weaker form of differentiability could be used (distribution theory for example).

The rest of the paper is organized as follows. Section 2 provides the necessary theoretical background of transport theory. Section 3 gives a brief overview of the continuous adjoint method in optimization. Section 4 presents the derivation of the *adjoint equation* in Transport Theory setting for radiance and its adjoint for the computation of the gradient of the cost function. Section 5 provides details of a simple implementation while Section 6 discusses several applications. Finally, we conclude in Section 7 and mention directions for future research.

But first as an appetizer we present some necessary results from functional analysis and fix notations.

## 1.1 Some Functional analysis

Let $\mathcal{F} = \mathcal{F}(\Omega, \mathbb{R}^n)$ be the **Hilbert** space of all functions mapping a continuous domain $\Omega$ to $\mathbb{R}^n$ equipped with the following *inner product*:

$$\langle f, g \rangle = \int_\Omega f^*(x) g(x) dx, \qquad \text{where } f, g \in \mathcal{F}.$$

This induces a norm on the space: $|f| = \sqrt{\langle f, f \rangle}$, we will also denote $\langle f \rangle = \langle 1, f \rangle = \int_\Omega f(x) dx$, the integral of $f$ over the entire domain $\Omega$. An *operator* is simply a linear function $\mathbf{A} : \mathcal{F} \to \mathcal{F}$. The *adjoint* of $\mathbf{A}$ is an operator denoted by $\mathbf{A}^*$ that satisfies

$$\langle \mathbf{A}f, g \rangle = \langle f, \mathbf{A}^* g \rangle \qquad \text{for all } f, g \in \mathcal{F}.$$

An operator that satisfies $\mathbf{A} = \mathbf{A}^*$ is called *self adjoint*. An important example is an operator defined by a kernel $K(x, y)$ as follows:

$$\mathbf{K}_x = \langle K(x,\cdot),\cdot \rangle = \int_\Omega K(x,y)g(y)dy.$$

We have the identities: $(\mathbf{A} + \mathbf{B})^* = \mathbf{A}^* + \mathbf{B}^*$ and $(\mathbf{AB})^* = \mathbf{B}^*\mathbf{A}^*$. Differentials of operators are to be understood as a **Fréchet Derivative**. The operator $\mathbf{A}$ is differentiable at $f$ if there exist an operator $\mathbf{D}$ (the derivative of $\mathbf{A}$ at $f$) such that

$$\lim_{|h|\to 0} \frac{|\mathbf{A}(f+h) - \mathbf{A}(f) - \mathbf{D}h|}{|h|} = 0.$$

For all sequences $h = \{h_n\}_{n=1}^\infty$ with $h_n \to 0$ as $n \to \infty$.

**Fun fact**: the Fréchet derivative of the Dirac delta operator: $\delta : C^\infty(\Omega) \to \mathbb{R} \subset C^\infty(\Omega) : \varphi \mapsto \varphi(0)$ is $\delta' : \varphi \mapsto -\varphi'(0)$. In general $\delta^{(k)} : \varphi \mapsto (-1)^k \varphi^{(k)}(0)$. It's just integration by parts via Riesz' Theorem. Yeah, a Dirac delta is not a weird function but an operator also known as a distribution.

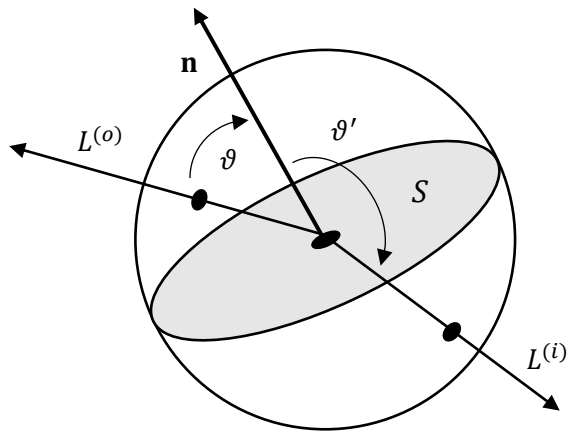

**Figure 1:** Geometry at a surface point and the decomposition of the radiance into outgoing and incoming parts.

## 2. Light Transport and the Adjoint Formulation[2]

We assume that our environment is comprised of a set of surfaces denoted by $S$. The environment between the surface is empty (no participating media) and light travels in straight lines between surface points. The properties of light like radiance are constant along each ray with changes occurring only at the surfaces. Consequently, the functions we will consider are defined over the space of rays spanned by the surfaces:

---

[2] This Section is inspired by Eric Veach's excellent PhD thesis [5].

$$\mathcal{R} = \{\bar{x} = x \to x' \text{ with } x, x' \in S\}.$$

This space is four-dimensional since each ray is defined by a the two-coordinates of its endpoints on each surface. We distinguish one of these points as the origin of the ray. This is indicated by the arrow notation: $\bar{x} = x \to x'$. Such that the ray with the opposite direction is denoted by $(-\bar{x}) = x' \to x$. The fundamental quantity in light transport is the *radiance* field:

$$L(\bar{x}) = L(x \to x')$$

which has physical units of radiant energy per area per solid angle: $W \cdot m^{-2} \cdot sr^{-1}$. In the following it will be convenient to distinguish between *incoming* radiances $L^{(i)}$ and *outgoing* radiances $L^{(o)}$ with respect to the normal **n** at a point on the surface. This is illustrated in Fig. 1. We have by convention that:

$$L(\bar{x}) = \begin{cases} L^{(o)}(\bar{x}) & \text{if } \cos\vartheta > 0 \\ L^{(i)}(-\bar{x}) & \text{if } \cos\vartheta < 0 \end{cases}$$

Where $\theta$ is the angle between the ray and the surface normal. These two fields are related by a *propagator* operator as follows:

$$(\mathbf{P}L)(\bar{x}) = L(-\bar{x})$$

It follows that we have $L^{(i)} = \mathbf{P}L^{(o)}$ and $L^{(o)} = \mathbf{P}L^{(i)}$. This operator is self-adjoint.

Light sources are modeled by an *emitter* field $L_0^{(o)}(\bar{x})$, while the interaction at the surfaces is given by a *scattering kernel*

$$K(\bar{x}, \bar{y}) = f_s(x \to x', y \to y')\delta(y - x')$$

Where $f_s$ is the *bi-directional scattering function (BSDF)* and $\delta$ is the Dirac-delta operator. The transport equation relates the radiance along a ray to sources and scattered radiances:

$$L(\bar{x}) = L_0^{(o)}(\bar{x}) + \int_\mathcal{R} K(\bar{x}, \bar{y})L(\bar{y})d\mu(\bar{y}) \quad (1)$$

Where the integration measure is defined by

$$d\mu(\bar{y}) = V(\bar{y})\frac{\cos\vartheta \cos\vartheta'}{|y - y'|^2}dydy'.$$

The visibility function $V(\bar{y})$ is equal to one when $y$ is visible from $y'$ along the ray $\bar{y} = y \to y'$ and zero otherwise (possibly smoothed for the sake of differentiability). Furthermore $\vartheta$ and $\vartheta'$ are the angles between the incoming and outgoing rays and the normal at the surface point $y = x'$. Eq. 1 can be written more compactly using a *transport* operator $\mathbf{T}(\bar{x}) = \langle K(\bar{x},\cdot),\cdot\rangle$ for the scattering operation:

$$L = L_0^{(o)} + \mathbf{T}L. \quad (2)$$

This is the ***transport equation*** for the radiance. Formally we can solve this equation using a Neumann[3] series as follows:

$$L = \mathbf{S}L_0^{(o)} = (\mathbf{I} - \mathbf{T})^{-1}L_0^{(o)} = (\mathbf{I} + \mathbf{T} + \mathbf{T}^2 + \mathbf{T}^3 + \cdots)L_0^{(o)}. \quad (3)$$

This series has a physical interpretation. The final radiance is equal to successive contributions involving increasing orders of scatter events. Given the radiance function $L$ we can measure its value using a ***receiver*** function $W_1^{(i)}(\bar{y})$ as follows:

$$I = I(L) = \int_{\mathcal{R}} W_1^{(i)}(\bar{y})L^{(i)}(\bar{y})d\mu(\bar{y}) = \langle W_1^{(i)}, L^{(i)} \rangle. \quad (4)$$

This is the quantity that we are essentially interested in computing. Using the propagator and the scattering operators we can rewrite the measured radiance:

$$I = \langle W_1^{(i)}, L^{(i)} \rangle = \langle W_1^{(i)}, \mathbf{PS}L_0^{(o)} \rangle = \langle (\mathbf{PS})^* W_1^{(i)}, L_0^{(o)} \rangle = \langle \mathbf{S}^* \mathbf{P} W_1^{(i)}, L_0^{(o)} \rangle = \langle W^{(o)}, L_0^{(o)} \rangle$$

The final radiance at a point can therefore be computed in two different ways. Either by propagating the source emitter or by propagating the receiving detector. For the latter we need to compute the adjoint of the scattering operator to obtain the ***importance*** field:

$$W = \mathbf{S}^* W_1^{(o)}. \quad (5)$$

From Eq. 3 it follows that:

$$\mathbf{S}^* = \mathbf{I} + \mathbf{T}^* + \mathbf{T}^{*2} + \mathbf{T}^{*3} + \cdots = (\mathbf{I} - \mathbf{T}^*)^{-1}.$$

Where $\mathbf{T}^* = \langle K^*(\bar{x}, \cdot), \cdot \rangle$ and $K^*(\bar{x}, \bar{y}) = K(-\bar{y}, -\bar{x})$. So that the importance field satisfies the ***adjoint transport equation***

$$W = W_1^{(o)} + \mathbf{T}^* W. \quad (6)$$

To summarize: we have two alternative ways to compute the radiance $I$ at the receiver/emitter. The first approach which we call the ***forward method*** is to solve for the radiance using the successive scatterings of the emitter (Eq. 3) and then compute the final radiance from

$$I = \langle W_1^{(i)}, \mathbf{S}L_0^{(i)} \rangle = \langle W_1^{(i)}, L_1^{(i)} \rangle.$$

In the forward mode we basically propagate radiance rays from the emitter to the receiver. Alternatively, we can use a ***backward method*** which solves for the importance by scattering the receiver (Eq. 5) and then compute the final radiance using

---

[3] Not named after the famous John Von Neumann (Hungarian-American) but Carl Neumann (German). In high school you probably learned that $1 + x + x^2 + \cdots = 1/(1-x)$ for $|x| < 1$. Well it is also true for operators when $|\mathbf{T}| < 1$. But in some sense $1 + 2 + 2^2 + \cdots = -1$ is fun nonsense.

$$I = \langle \mathbf{S}^* W_1^{(o)}, L_0^{(o)} \rangle = \langle W_0^{(o)}, L_0^{(o)} \rangle.$$

The backward mode traces importance rays from the receiver to the emitter. Hybrid schemes are also possible. Where one can start two sets of propagating rays, one starting from the emitter and the other starting at the receiver connecting them somewhere in the middle. This technique is known as **bi-directional ray tracing** in computer graphics.

This concludes our brief overview of light transport theory and the role of the adjoint transport operator in connecting radiance and importance.

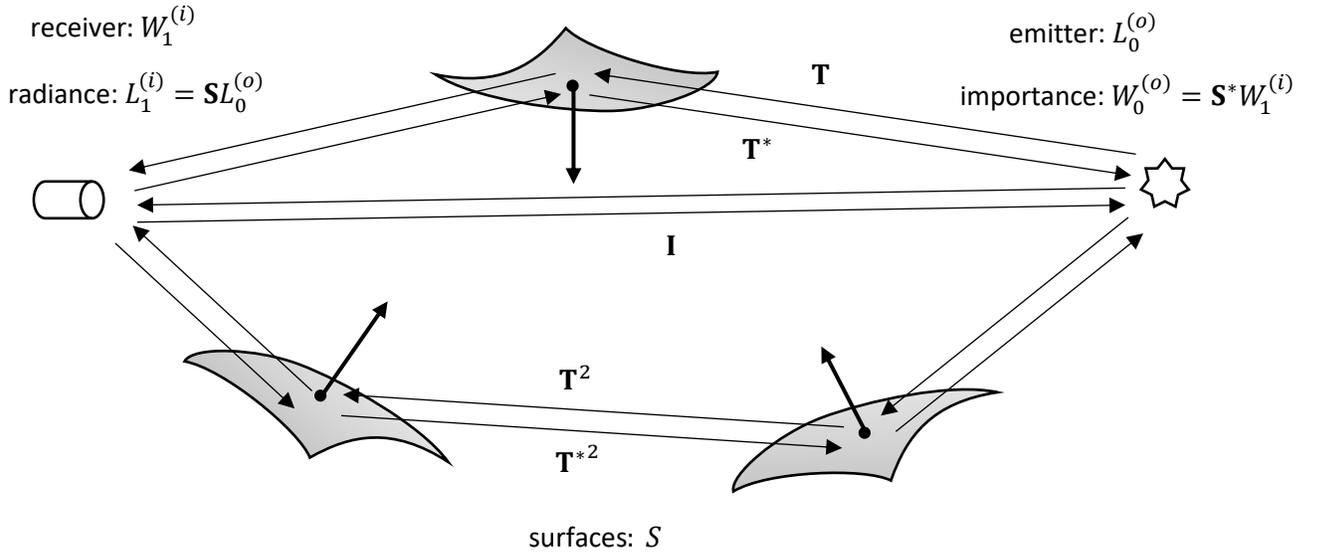

**Figure 2:** Radiance (forward) and Importance (backward) propagation.

## 3. The Continuous Adjoint Method in Optimization[4]

The goal of optimization is to find the minimum (maximum) of a *cost function* $\mathcal{J}(u, \theta)$ depending on a **state** $u$ and a **control** $\theta$. Both the state and the control are continuous functions depending on a variable $\omega \in \Omega \subset \mathbb{R}^d$. The state and the control are also constrained to satisfy an equation $E(u, \theta) = 0$. For example, in the case of Ordinary Differential Equations, the continuous variable is time and the state must satisfy a differential equation: $E(u, \theta) = -\dot{u}(t) + f(u(t), \theta(t))$. The fundamental problem of continuous optimization (and machine learning) can be stated concisely as:

$$\textbf{Find } \theta^* = \underset{\theta}{\mathrm{argmin}}\, \mathcal{J}(u, \theta) \textbf{ such that } E(u, \theta) = 0, \quad (7)$$

---

[4] The continuous adjoint method was first introduced by Pontryagin and coworkers in [4]. The article by Giles and Pierce is a very good introduction [1]. The adjoint method was first applied in computer graphics to control fluid-like animations [3].

where $u(\omega) \in \mathbb{R}^n$, $\theta(\omega) \in \mathbb{R}^m$ and $E(u, \theta) \in \mathbb{R}^k$. We assume that the cost function is defined over the entire domain:

$$\mathcal{J}(u, \theta) = \int_\Omega J\big(u(\omega), \theta(\omega)\big) d\omega = \langle J(u, \theta) \rangle.$$

However, in many applications the cost function is only defined at a finite set of points $\widehat{\omega}_s \in \Omega$:

$$\mathcal{J}(u, \theta) = \frac{1}{2} \sum_{s=1}^{N} |u(\widehat{\omega}_s) - \hat{u}_s|^2$$

Where the $\hat{u}_s \in \mathbb{R}^n$ ($s = 1, \cdots, N$) are the desired states. This is a common type of cost function for least square optimization and supervised (deep) learning.

In a smooth setting where all functions are assumed to de differentiable optimization and learning algorithms rely heavily on the gradient of the cost function. Consequently, a lot of research in these fields is devoted to computing this gradient. In fact, it is **the** fundamental challenge. The research described in this paper is no exception! For example, both gradient descent and quasi-Newton iterative methods rely heavily on a gradient of the cost function.

More precisely, given a cost function we are interested in computing the gradient of the cost function with respect to the controls:

$$\delta \mathcal{J} = \frac{d\mathcal{J}}{d\theta}.$$

That is the *holy grail* we are after. Notice that the "$\delta$" symbol is a short-hand for "$d/d\theta$" not the Dirac-delta function: $\delta X$ means a variation of $X$ with respect to the control $\theta$.

Constrained optimization problems like Eq. 7 can be transformed into unconstrained problems using the machinery of Lagrange multipliers. In the continuous setting one introduces a ***Lagrange multiplier*** function $p(\omega)$. We then augment the cost function with a penalty term involving the multiplier and the constraint:

$$\mathcal{L}(u, p, \theta) = \mathcal{J}(u, \theta) + \int_\Omega p(\omega)^* E\big(u(\omega), \theta(\omega)\big) d\omega = \langle J(u, \theta) \rangle + \langle p, E(u, \theta) \rangle.$$

This is the less familiar continuous version of the Lagrangian. The necessary conditions for optimality are (where the derivatives are Fréchet):

$$\frac{\partial \mathcal{L}}{\partial u} = 0 \quad \frac{\partial \mathcal{L}}{\partial p} = 0 \quad \text{and} \quad \frac{\partial \mathcal{L}}{\partial \theta} = 0.$$

From the first condition we get an adjoint equation for the multiplier (see Appendix A)

$$\left(\frac{\partial E}{\partial u}\right)^* p = -\frac{\partial J}{\partial u}. \quad (8)$$

This equation is independent of the controls! This is the key reason why the adjoint method is so popular in optimization and machine learning. The consequence is that computing the gradient is no more costly then computing the function itself. The Lagrange multiplier is usually called the **adjoint function** in the optimization literature. Intuitively, the adjoint function models the sensitivity of the cost function with respect to the state independently of the controls. Once the adjoint function is computed we obtain the gradient of the cost function with respect to the controls as follows (see Appendix A)

$$\frac{dJ}{d\theta} = \langle p, \frac{\partial E}{\partial \theta}\rangle + \langle \frac{\partial J}{\partial \theta}\rangle. \quad (9)$$

Computing the gradient of the cost function therefore involves two steps. The solution of the adjoint equation for $p(\omega)$ and the evaluation of the gradient. These equations are very general and can be applied to most optimization and machine learning problems. Next, we apply this methodology to the transport theory of light propagation.

## 4. Adjoint Method Applied to Transport Theory

We now combine the adjoint method with the transport equations. An example of a cost function in rendering is

$$J(L, \theta) = \frac{1}{2}\sum_{s=1}^{N_s}|I_s(L) - \hat{I}_s|^2 + \frac{1}{2}\varepsilon|\theta|^2. \quad (10)$$

Where the sum is over the receivers and the $\hat{I}_s$ are some target values and $\varepsilon \geq 0$ models the smoothness of the control. In this case the gradient is:

$$\frac{dJ}{d\theta} = \sum_{s=1}^{N_s}(I_s(L) - \hat{I}_s)\frac{dI_s}{d\theta} + \varepsilon\theta.$$

Our method can of course handle more general cost functions. But it is helpful to hold this typical example in your mind. Why? Because we are really after computing the gradient of the radiance with respect to the controls. Let that sink in. From Eq. 4 we have that

$$\frac{dI_s}{d\theta} = \langle \frac{dW_1^{(i)}}{d\theta}, L^{(i)}\rangle + \langle W_1^{(i)}, \frac{dL^{(i)}}{d\theta}\rangle.$$

In general, we assume that the transport operator and the emitters depend on the control function $\theta(\omega)$. Consequently, our transport equation (Eq. 2) becomes:

$$E(L, \theta) = -L + \mathbf{T}_\theta L + L_{0,\theta}^{(o)} = 0.$$

Where the subscript denotes dependence of a function/operator on the control $\theta$ **not** differentiation. Its differential with respect to the radiance is:

$$\frac{\partial E}{\partial L} = -\mathbf{I} + \mathbf{T}_\theta.$$

And an equation for the adjoint function $p(\omega)$ follows from Eq. 8:

$$-p\mathbf{I} + (\mathbf{T}_\theta)^* p = -\frac{\partial J}{\partial L}. \quad (11)$$

This equation can be written using the Neumann series (Eq. 5)

$$p_0 = (\mathbf{S}_\theta)^* p_1. \quad (12)$$

*Equation 12 is the main result of this paper.*

This equation is exactly the adjoint transport equation for the importance field with a different source term:

$$p_1 = \frac{\partial J}{\partial L}. \quad (13)$$

And we have for the particular cost function given by Eq. 10 that

$$p_1 = (I_s(L) - \hat{I}_s)\frac{\partial I_s(L)}{\partial L}.$$

Eq. 12 does not depend on the number of controls and is therefore as efficient to solve as the adjoint transport equation for the importance. It also does not need the computation of derivatives with respect to the controls. This is a direct consequence of the fact that the transport operator is linear with respect to the radiance. The same operator is used for the equation of the adjoint function $p(\omega)$. While solving the adjoint through propagation we compute the gradients of the cost function sequentially with respect to the controls at each scatter/emitter from Eq. 9:

$$\frac{dJ}{d\theta} = \langle p, \left(\frac{\partial \mathbf{T}_\theta}{\partial \theta}\right) L + \frac{\partial L_{0,\theta}^{(o)}}{\partial \theta} \rangle + \frac{\partial J}{\partial \theta}. \quad (14)$$

*Equation 14 is the second main result of this paper.*

This step requires the derivatives of the transport operator and the emitter with respect to the controls. These derivatives can be computed analytically or through automatic differentiation.

## 5. Implementation

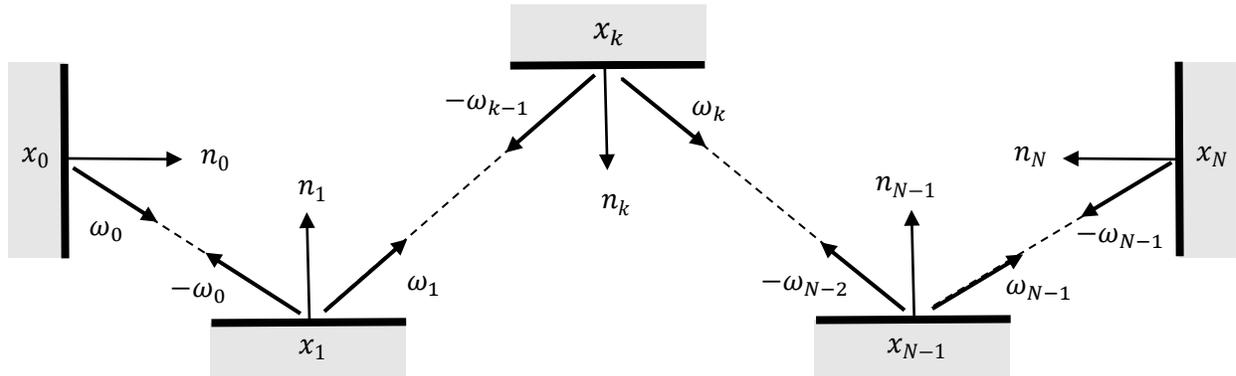

**Figure 3:** Geometry of a path.

### 5.1 Path Tracing

First, we introduce the standard notations used in computer graphics: each ray is defined by a position $x$ and a direction $\omega$. Let $n$ be the surface normal at this point. The outgoing radiance $L(\omega_{out})$ due to an incoming radiance $L(\omega_{in})$ is given by a **B**idirectional **R**eflectance **D**istribution **F**unction (**BRDF**):

$$L(\omega_{out}) = \rho(n, -\omega_{in}, \omega_{out})(n \cdot \omega_{in}) L(\omega_{in}).$$

Where $\rho(n, \omega_{in}, \omega_{out})$ is the BRDF. The minus sign in front of the incoming direction $\omega_{in}$ is there because the BDRF is usually defined with both vectors pointing outwards from the surface. However, in path tracing the directions shown in Figure 4 are more natural. Similarly, the outgoing adjoint $P(\widehat{\omega}_{out})$ in direction $\widehat{\omega}_{out}$ due to an incoming adjoint $P(\widehat{\omega}_{in})$ from a direction $\widehat{\omega}_{in}$ is related via the adjoint $\rho^*$ of the BRDF:

$$P(\widehat{\omega}_{out}) = \rho^*(n, -\widehat{\omega}_{in}, \widehat{\omega}_{out})(n \cdot \widehat{\omega}_{in}) P(\widehat{\omega}_{in}).$$

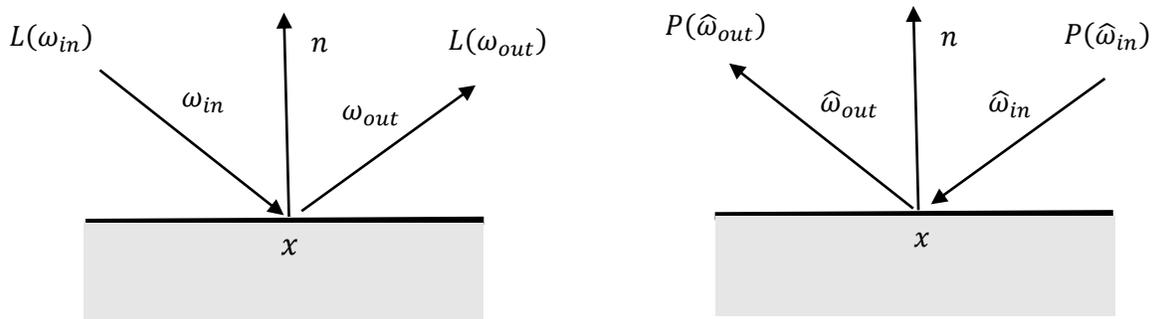

**Figure 4:** Definitions of the BRDF, the radiance and its adjoint.

The situation is shown in the Figure 4. The angles for the radiance in a forward pass are related to the angles for the importance by the following relations: $\widehat{\omega}_{out} = -\omega_{in}$ and $\widehat{\omega}_{in} = -\omega_{out}$.

We can generate random paths in the environment as follows. We start at the receiver and then trace rays until we hit an emitter as shown in Figure 3. For each incoming ray we generate an outgoing ray randomly from a **P**robability **D**istribution **F**unction (**PDF**) $p(\omega)$. Appendix B shows how to generate such rays. The random path so created is denoted by a sequence of points: $x_0, x_1, \cdots, x_N$ with associated surface normals $n_0, n_1, \cdots, n_N$ as shown in Figure 5 above. The receiver is situated at the start point $x_0$ and the emitter is located at the end point $x_N$. For each pair of points on the path we associate a unit direction:

$$\omega_k = \frac{x_{k+1} - x_k}{\|x_{k+1} - x_k\|}, \quad k = 0, \cdots, N-1.$$

We have by construction that $n_k \cdot \omega_k > 0$ and $n_k \cdot \omega_{k-1} < 0$. Since radiances and adjoints are constant along each ray we have that:

$$L(\omega_k) = L(-\omega_k) \text{ and } P(\omega_k) = P(-\omega_k).$$

Our goal is to minimize a cost function which we assume for simplicity to only depend on the radiance at the emitter and a set of control variables $\theta$:

$$J \leftarrow J(L(\omega_0), \theta).$$

Our goal is to compute the gradient $\frac{dJ}{d\theta}$ using the adjoint method described above.

We compute the radiance at the receiver in a forward pass. First, we initialize the radiance at the emitter:

$$L(\omega_{N-1}) = L_{N,\theta}(-\omega_{N-1}).$$

Where the $\theta$ subscript indicates that a function depends on the controls. The radiances along the path are then computed using the BRDF $\rho_{k,\theta}$ and the PDF $p_{k,\theta}$ of the surfaces:

$$L(\omega_{k-1}) = \rho_{k,\theta}(n_k, \omega_k, -\omega_{k-1})(n_k \cdot \omega_k)L(\omega_k)/p_{k,\theta}(\omega_k) \text{ for } k = N-1, \cdots, 1.$$

This gives us the incoming radiance $L(\omega_0)$ at the receiver. From this radiance we can compute the cost function $J(L(\omega_0), \theta)$ and its adjoint

$$P(\omega_0) = \frac{\partial J}{\partial L}(L(\omega_0), \theta).$$

Then we proceed with a backward pass to compute the adjoints starting with $P(\omega_0)$:

$$P(\omega_k) = \rho^*_{k,\theta}(n_k, -\omega_{k-1}, \omega_k)(-n_k \cdot \omega_{k-1})P(\omega_{k-1})/p_{k,\theta}(\omega_{k-1}) \text{ for } k = 1, \cdots, N-1.$$

The gradient of the cost function $\frac{dJ}{d\theta}$ is computed alongside with the adjoint as follows. We initially set this gradient to zero

$$\frac{dJ}{d\theta} = 0$$

and then incrementally updated it at each step of the backward pass:

$$\frac{dJ}{d\theta} \mathrel{+}= \left(P(\omega_{k-1}) \cdot L(\omega_k)\right) \frac{\partial \rho_{k,\theta}}{\partial \theta}.$$

Although this seems like an expensive step when there are many controls. We notice that in general only the subset of controls that the BRDF $\rho_{k,\theta}$ depends on must be updated. In practice this subset is much smaller than the total number of controls: the entire control vector is rarely updated. We finish the backward trace with an update of the gradient at the emitter:

$$\frac{dJ}{d\theta} \mathrel{+}= \left(P(\omega_{N-1}) \cdot \frac{\partial L_{N,\theta}}{\partial \theta}\right).$$

The pair $\left(J, \frac{dJ}{d\theta}\right)$ can then be fed to an optimizer to return a new set of controls $\theta$ and the whole process is iterated.

**5.2 A Simple Example**[5]

As a simple example we consider the diffuse Lambertian BRDF:

$$\rho_\theta(n, \omega_0, \omega_1) = \frac{\theta_1}{\pi},$$

a simple emitter:

$$L_\theta(\omega) = \theta_2 E_N$$

and a PDF that generates cosine weighted random samples:

$$p(\omega) = \frac{\cos \vartheta}{\pi} \quad \text{with} \quad \omega = (\vartheta, \varphi).$$

The control is therefore a 2-vector: $\theta = (\theta_1, \theta_2)$. We assume a very simple cost function:

$$J(L) = \frac{1}{2}\left(L - \hat{L}\right)^2,$$

where $\hat{L}$ is the desired output for the output radiance. The adjoint at the receiver is then given by:

---

[5] See also Appendix C for details on the Cook-Torrance BRDF.

$$\frac{\partial J}{\partial L} = L - \hat{L}.$$

The derivatives of the BRDF divided by the PDF and of the emission function are:

$$\frac{\partial(\rho_\theta/p)}{\partial \theta} = \begin{pmatrix} 1 \\ 0 \end{pmatrix} \text{ and } \frac{\partial L_\theta}{\partial \theta} = \begin{pmatrix} 0 \\ 1 \end{pmatrix} E_N.$$

This results in the following pseudo-code. Notice that we have $\frac{(n_k \cdot \omega_k)}{p(\omega_k)} = \pi$ and $\frac{-(n_k \cdot \omega_{k-1})}{p(\omega_{k-1})} = \pi$ because our random directions are assumed to be cosine weighted. This leads to the following simple path tracing algorithm.

**Forward pass**:

$$L_{N-1} = \theta_2 E_N$$

$$\text{for } k = N-1, \cdots, 1 \text{ do } \{L_{k-1} = \theta_1 L_k$$

**Backward pass**:

$$\frac{dJ}{d\theta} = \begin{pmatrix} 0 \\ 0 \end{pmatrix}$$

$$P_0 = L_0 - \hat{L}$$

$$\text{for } k = 1, \cdots, N-1 \text{ do } \begin{cases} P_k = \theta_1 P_{k-1} \\ \frac{dJ}{d\theta_1} \mathrel{+}= (P_{k-1} \cdot L_k) \end{cases}$$

$$\frac{dJ}{d\theta_2} \mathrel{+}= (P_{N-1} \cdot E_N)$$

It doesn't get simpler than this!

## 6. Results

To validate our model, we have implemented a simple "Vanilla" text-book style path tracer. An outline of the implementation in pseudo code is given in Appendix D. We decided to implement our own path tracer rather than modifying an existing one so that we could focus on the core algorithm not on compatibility and build issues. It also offers more flexibility in debugging and visualizing our results. Our path tracer only handles ray/sphere and ray/quad intersections. Consequently, our scenes are restricted to quads and spheres. Also, we consider only three types of materials.

**M1**: A pure emitter defined by its **emission strength** ($\theta_1$).

**M2**: A specular Phong-Blinn BRDF / PDF modelled by an **ambient term** ($\theta_2$), a **diffuse term** ($\theta_3$), a **specular term** ($\theta_4$) and finally an **exponent** ($\theta_5$).

**M3**: A diffuse BRDF / PDF defined by an **ambient term** ($\theta_6$) and a **diffuse term** ($\theta_7$).

The associated controls are given in parenthesis, there are seven of them which we represent by a vector $\theta = (\theta_1, \theta_2, \theta_3, \theta_4, \theta_5, \theta_6, \theta_7)$. Our goal is to compute the gradient of the cost function with respect to these controls.

To validate our model, we consider a simple "Cornell Box"-type scene. The scene is comprised of a box with six quad walls having material **M3**, one quad light source at the top with material **M1** and one sphere in the center with material **M2**. The receiver is a pinhole camera defined by a screen located inside the box. Figure 5 depicts the scene in our visualizer.

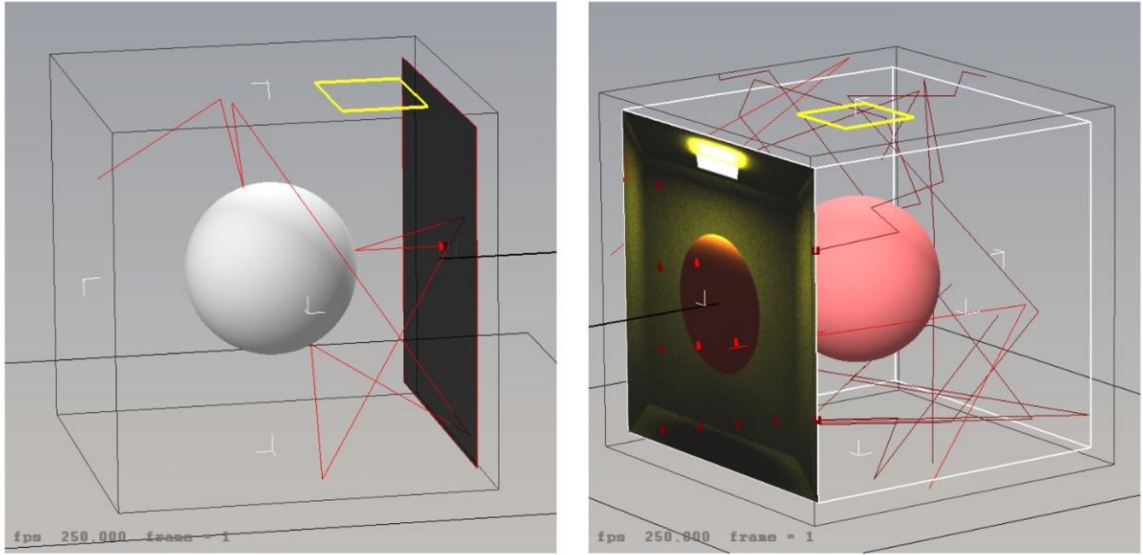

**Figure 5.** Two snapshots of our simple Path tracer. Left visualization of a single path and right a view of a rendered image and some selected paths.

Our visualizer can depict any quantity we like for debugging purposes. On the left of Figure 5 we show a single path, while on the right we show a subset of paths and the resulting rendered image. The ability to trace a single path was very useful in comparing our model with estimations obtained from a finite difference approximation of the gradients with respect to the controls. The approximation is obtained by choosing a small number $\varepsilon$ and computing

$$\frac{dJ}{d\theta_k} \cong \frac{J(\theta_1, \cdots, \theta_k + \varepsilon, \cdots, \theta_K) - J(\theta_1, \cdots, \theta_k - \varepsilon, \cdots, \theta_K)}{2\varepsilon}$$

for each control. Therefore, we must run the path tracer a number $2K$ of times where $K = 7$ in our example. The right choice for the value of the perturbation $\varepsilon$ is tricky. When $\varepsilon$ is too big the estimate is inaccurate and when it is too small, we loose numerical precision because we are subtracting two

nearby numbers. This is another strong argument for computing gradients using adjoints that do not

```
J = 3.3162e-06
dJ/dcontrol =
0 0 3.56703 22.1649 0.134605 10 13.3333
---------- Numerical gradients ---------
EPS = 0.1                                   EPS = 0.0001                                EPS = 1e-07                                 EPS = 1e-10
numerical dJ/dcost:                         numerical dJ/dcost:                         numerical dJ/dcost:                         numerical dJ/dcost:
( 1.000000 - 1.000000 ) / 2*EPS = 0.000000  ( 1.000000 - 1.000000 ) / 2*EPS = 0.000000  ( 1.000000 - 1.000000 ) / 2*EPS = 0.000000  ( 1.000000 - 1.000000 ) / 2*EPS = 0.000000
( 1.000000 - 1.000000 ) / 2*EPS = 0.000000  ( 1.000000 - 1.000000 ) / 2*EPS = 0.000000  ( 1.000000 - 1.000000 ) / 2*EPS = 0.000000  ( 1.000000 - 1.000000 ) / 2*EPS = 0.000000
( 1.411622 - 0.691398 ) / 2*EPS = 3.601122  ( 1.000357 - 0.999644 ) / 2*EPS = 3.565848  ( 1.000001 - 1.000000 ) / 2*EPS = 4.768371  ( 1.000000 - 1.000000 ) / 2*EPS = 0.000000
( 6.075682 - 0.053732 ) / 2*EPS = 30.109749 ( 1.002218 - 0.997785 ) / 2*EPS = 22.164581 ( 1.000002 - 0.999998 ) / 2*EPS = 19.669531 ( 1.000000 - 1.000000 ) / 2*EPS = 0.000000
( 1.013496 - 0.986575 ) / 2*EPS = 0.134604  ( 1.000014 - 0.999987 ) / 2*EPS = 0.134408  ( 1.000000 - 1.000000 ) / 2*EPS = 0.000000  ( 1.000000 - 1.000000 ) / 2*EPS = 0.000000
( 2.250000 - 0.250000 ) / 2*EPS = 10.000000 ( 1.001000 - 0.999000 ) / 2*EPS = 10.000169 ( 1.000001 - 0.999999 ) / 2*EPS = 8.642673  ( 1.000000 - 1.000000 ) / 2*EPS = 0.000000
( 3.160494 - 0.197531 ) / 2*EPS = 14.814816 ( 1.001334 - 0.998668 ) / 2*EPS = 13.332068 ( 1.000001 - 0.999999 ) / 2*EPS = 13.411044 ( 1.000000 - 1.000000 ) / 2*EPS = 0.000000
```

**Figure 6.** Comparison of the gradients obtained with our method (top/left) and finite difference estimates for different values of $\varepsilon$.

depend on an arbitrary small parameter. In fact our adjoints are accurate up to machine precision. We found a sweet spot by varying the perturbation: $\varepsilon = 0.1, 0.0001, \ldots$ In Figure 6 we show these estimates for different $\varepsilon$ along with the results from our method at the top. We observe good agreement. Next, we show renderings of the final image (top left) and the gradient for each of the seven controls. Figure 7 computes paths without importance sampling while in Figure 8 the paths are computed with a Phong-Blinn importance sampling as described in Appendix B.

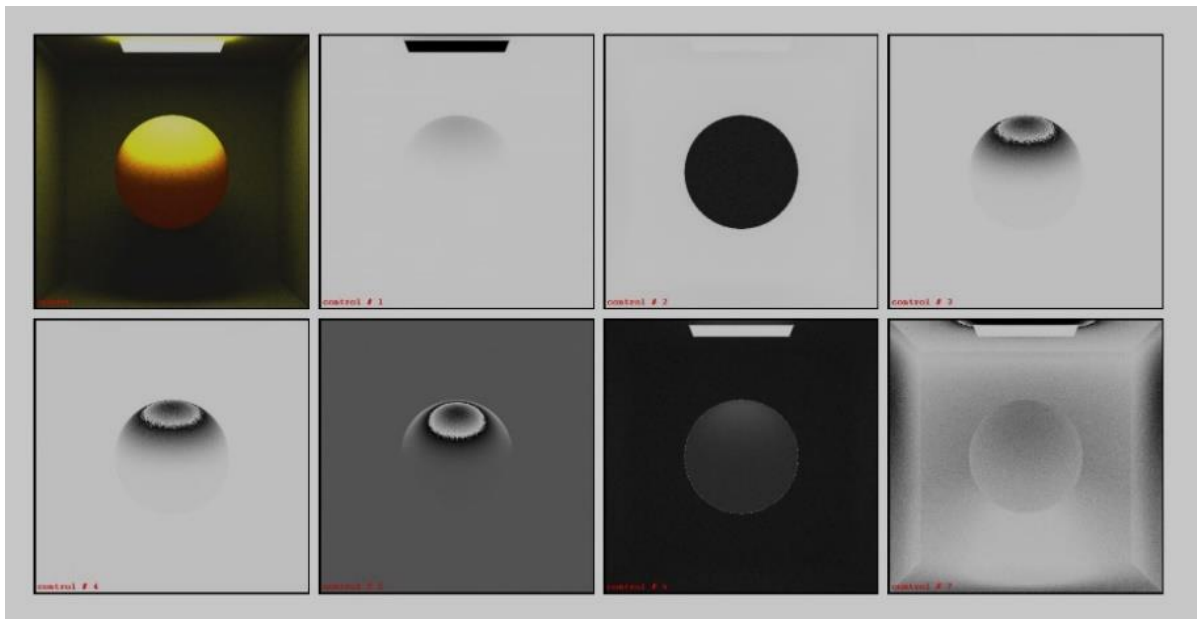

**Figure 7.** Renderings of the image (top/left) and the gradients with respect to the controls with cosine-weighted uniform sampling.

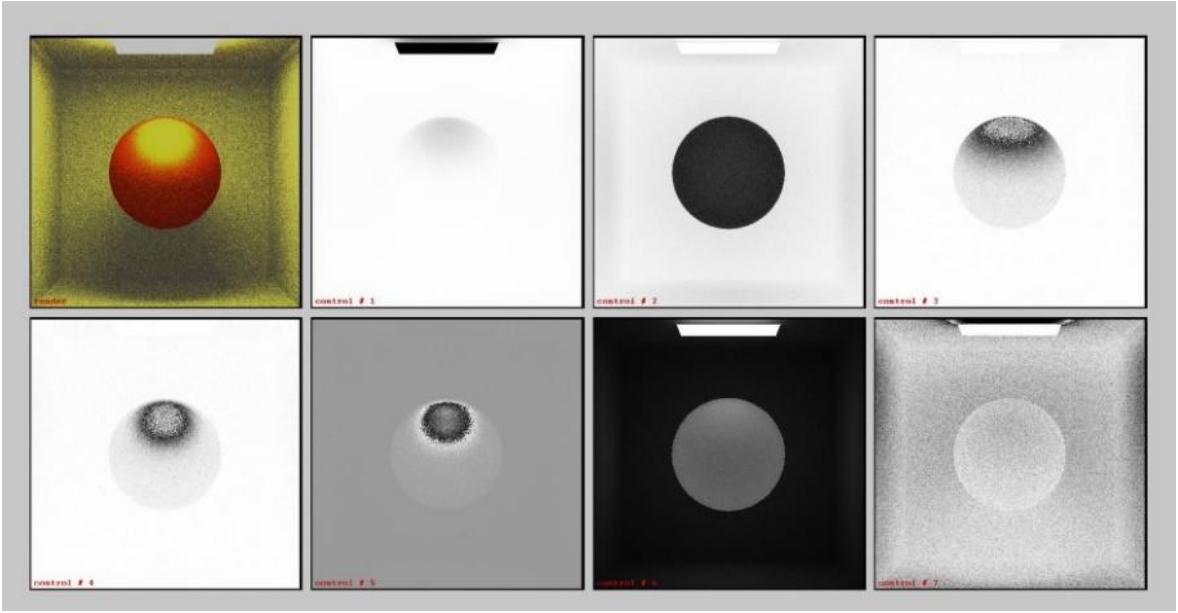

**Figure 8.** Renderings of the image (top/left) and the gradients with respect to the controls with Phong-Blinn importance sampling.

## 7. Conclusions and Future Work

In this work we have presented a novel general method to compute gradients of radiance in the context of transport theory. To achieve this, we have derived an equation for the adjoint/Jacobian of the cost function from the general theory of adjoints in optimization theory. We have shown that this can easily be implemented in a simple home brewed path tracer. The results show good agreement with a finite difference computation of the cost function.

In the future we want to extend the theory to more general settings including volumetric scatterers and perhaps include diffraction effects. In these cases, we have to deal with an integro-integral equation and a Kirchhoff integral, respectively. Also, we want to implement this method in various existing GPU-based path tracers like Fermat.

# Appendices

## A. Adjoint Equation

In this appendix we derive the adjoint equation (Eq. 8) and an expression for the derivative of the cost function with respect to the controls (Eq. 9). Recall that the augmented Lagrangian is defined as

$$\mathcal{L}(u, p, \theta) = \langle J(u, \theta) \rangle + \langle p, E(u, \theta) \rangle.$$

Stationarity with respect to the state implies

$$0 = \frac{\partial \mathcal{L}}{\partial u} \delta u = \langle \frac{\partial J}{\partial u}, \delta u \rangle + \langle p, \frac{\partial E}{\partial u} \delta u \rangle \quad \rightarrow \quad \langle \frac{\partial J}{\partial u}, \delta u \rangle = - \langle p, \frac{\partial E}{\partial u} \delta u \rangle. \quad (A.1)$$

From the same equation and using the definition of the adjoint we have that

$$0 = \langle \frac{\partial J}{\partial u}, \delta u \rangle + \langle p, \frac{\partial E}{\partial u} \delta u \rangle = \langle \frac{\partial J}{\partial u}, \delta u \rangle + \langle \left(\frac{\partial E}{\partial u}\right)^* p, \delta u \rangle = \langle \frac{\partial J}{\partial u} + \left(\frac{\partial E}{\partial u}\right)^* p, \delta u \rangle.$$

Since this equation must hold for all functions $\delta u$ we get Eq. 8:

$$\left(\frac{\partial E}{\partial u}\right)^* p = -\frac{\partial J}{\partial u}.$$

Now consider the second condition:

$$0 = \frac{\partial \mathcal{L}}{\partial p} = E(u, \theta).$$

This is simply the equation that our state must satisfy the constraint and implies that its differential vanishes

$$0 = \delta E = \frac{\partial E}{\partial u} \delta u + \frac{\partial E}{\partial \theta} \quad \rightarrow \quad \frac{\partial E}{\partial u} \delta u = -\frac{\partial E}{\partial \theta}. \quad (A.2)$$

With these results we can compute the gradient of the cost function

$$\delta J = \langle \frac{\partial J}{\partial u}, \delta u \rangle + \langle \frac{\partial J}{\partial \theta} \rangle \underset{A.1}{=} - \langle p, \frac{\partial E}{\partial u} \delta u \rangle + \langle \frac{\partial J}{\partial \theta} \rangle \underset{A.2}{=} \langle p, \frac{\partial E}{\partial \theta} \rangle + \langle \frac{\partial J}{\partial \theta} \rangle.$$

This is Eq. 9.

## B. Sampling from a PDF

For simplicity we assume that the PDF is isotropic and thus only depends on the elevation angle:

$$p(\omega) = 2\pi \, p(\vartheta).$$

We then generate a random sample $\omega$ from this distribution using the *Cumulative Probability Distribution* (*CDF*):

$$P(\vartheta) = \text{Prob}(t \leq \vartheta) = 2\pi \int_0^\vartheta p(t) \sin t \, dt.$$

This is a mapping from elevation angles in $[0, \frac{\pi}{2}]$ to the unit interval $[0,1]$. Using the inverse of the CDF we can directly generate a $p$-distributed random angle $\vartheta$ from a uniformly distributed $u \sim U(0,1)$:

$$\vartheta = P^{-1}(u).$$

As an example, consider the cosine-weighted Phong-Blinn PDF depending on an exponent parameter $\alpha \geq 0$ (for $\alpha = 0$ one obtains a cosine weighted Lambertian PDF). This PDF handles all cases considered in this paper:

$$p_\alpha(\vartheta) = \frac{\alpha + 2}{2\pi} (\cos \vartheta)^{\alpha+1}.$$

This gives the following CDF:

$$P_\alpha(\vartheta) = (\alpha + 2) \int_0^\vartheta (\cos t)^{\alpha+1} \sin t \, dt = 1 - (\cos \vartheta)^{\alpha+2}.$$

With an inverse equal to:

$$\vartheta = \Phi_\alpha(u) = P_\alpha^{-1}(u) = \text{acos}\left((1-u)^{\frac{1}{\alpha+2}}\right).$$

These results give us a recipe to generate random vectors for our path tracer from the PDFs given above. The general algorithm works as follows. To generate a random vector $\omega$ in a world coordinate frame $(X, Y, Z)$ with the $Z$-vector being the normal direction:

**Generate**: $u_1, u_2 \sim U(0,1)$.

**Set**: $\vartheta = \Phi_\alpha(u_1)$ and $\varphi = 2\pi u_2$.

**Random vector**: $\bar{\omega} = (\bar{\omega}_x, \bar{\omega}_y, \bar{\omega}_z) = (\cos \varphi \cos \vartheta, \sin \varphi \cos \vartheta, \sin \vartheta)$.

**Convert to world space**: $\omega = \bar{\omega}_x X + \bar{\omega}_y Y + \bar{\omega}_z Z$.

## C. The Cook-Torrance and Phong-Blinn BRDF

In this appendix we give all the details for the Phong-Blinn model. We will use a variant of the Cook-Torrance BRDF to simplify some of the resulting expressions. The goal here is not to faithfully reproduce the original Cook-Torrance BRDF but to show the details on how to implement such a model in our framework. The Cook-Torrance BRDF is usually defined as follows:

$$\rho(n, \omega_i, \omega_o) = F(n, \omega_i, \omega_o) G(n, \omega_i, \omega_o) D(\omega_m),$$

Where $F$ account for Fresnel effects which we will set equal to one. $G$ is a geometric factor that accounts for shadowing and the spread of the solid angles. We use the following expression:

$$G(n, \omega_i, \omega_o) = \frac{(\omega_m \cdot n)}{4(\omega_m \cdot \omega_o)(\omega_i \cdot n)} = \frac{\cos \vartheta_m}{4 \cos \vartheta_{mo} \cos \vartheta_i}.$$

Where $\omega_m$ is the vector halfway between the incoming and the outgoing directions:

$$\omega_m = \frac{\omega_i + \omega_o}{\|\omega_i + \omega_o\|}.$$

This vector is the normal that perfectly reflects the incoming light into the outgoing direction and is distributed according to a **Normal Density Function** (**NDF**) $D(\omega_m)$. In this appendix we will consider the Phong-Blinn NDF ($\omega_m = (\vartheta_m, \varphi_m)$):[6]

$$D_{PB,\alpha}(\vartheta_m, \varphi_m) = \frac{\alpha + 2}{2\pi} (\cos \vartheta_m)^\alpha \sin \vartheta_m.$$

Notice that this distribution is not normalized since the right normalizing term should be $(\alpha + 1)/2\pi$. Since we are not sampling from this distribution but the one given below this does not matter. This choice does however simplify the expressions below. The incoming direction is a function of the normal and the outgoing direction through the reflection law:

$$\omega_i = R_o(\omega_m) = 2(\omega_m \cdot \omega_o)\omega_m - \omega_o.$$

And the corresponding differentials satisfy:

$$d\omega_i = dR_o(\omega_m) = 4(\omega_m \cdot \omega_o)d\omega_m = 4 \cos \vartheta_{mo} \sin \vartheta_m \, d\vartheta_m d\varphi_m.$$

The PDF for the cosine weighted Phong-Blinn model is given by:

$$p_\alpha(\vartheta_m, \varphi_m) = \frac{\alpha + 2}{2\pi} (\cos \vartheta_m)^{\alpha+1} \sin \vartheta_m.$$

---

[6]We point out that other choices for the NDF such as Beckmann or GGX are currently more popular. The same methodology described here applies to those models as well.

This distribution is normalized, and we can therefore use it to sample random normals using the procedure described in Appendix B.

In importance path tracing we evaluate the outgoing radiance as follows:

$$dL(\omega_o) = \frac{\rho_\alpha(n, \omega_i, \omega_o)}{p_\alpha(\omega_m)} (\omega_i \cdot n) L(\omega_i) d\omega_i.$$

We can rewrite this in terms of the angles $(\vartheta_m, \varphi_m)$ as follows where we also use all the expressions derived above:

$$dL(\vartheta_o, \varphi_o) = \frac{\rho_\alpha(\vartheta_i, \theta_o)}{p_\alpha(\vartheta_m, \varphi_m)} \cos \vartheta_i \, 4 \cos \vartheta_{mo} \sin \vartheta_m \, L(R_o(\omega_m)) d\vartheta_m d\varphi_m$$

$$= \frac{\left(\frac{\alpha+2}{2\pi}(\cos \vartheta_m)^\alpha \sin \vartheta_m \cos \vartheta_m \right)}{4 \cos \vartheta_{mo} \cos \vartheta_i} \times \cos \vartheta_i \, 4 \cos \vartheta_{mo} \sin \vartheta_m \, L(R_o(\omega_m)) d\vartheta_m d\varphi_m$$
$$\left(\frac{\alpha+2}{2\pi}(\cos \vartheta_m)^{\alpha+1} \sin \vartheta_m\right)$$

$$= L(R_o(\omega_m)) \sin \vartheta_m \, d\vartheta_m d\varphi_m.$$

After grand eliminations, we get the simple update rule:

$$dL(\omega_o) = L(R_o(\omega_m)) \sin \theta_m(\alpha) \, d\vartheta_m d\varphi_m.$$

To update the gradient of the cost function with respect to the exponent parameter $\alpha$ we need to differentiate this expression. The sine term depends on $\alpha$ through the sampling procedure (see Appendix B):

$$s(\alpha) = \sin \theta_m(\alpha) = \sqrt{1 - (1-u)^{\frac{2}{\alpha+2}}}.$$

And its derivative is:

$$\frac{ds}{d\alpha}(\alpha) = \frac{(1-u)^{\frac{2}{\alpha+2}} \log(1-u)}{(\alpha+2)^2 \sqrt{1-(1-u)^{\frac{2}{\alpha+2}}}} = \frac{\cos^2 \vartheta_m \log(\cos^{\alpha+2} \vartheta_m)}{(\alpha+2)^2 \sin \vartheta_m} = \frac{\cos^2 \vartheta_m \log(\cos \vartheta_m)}{(\alpha+2) \sin \vartheta_m}.$$

## D. A Simple Path Tracer

In this Appendix we describe our simple path tracer in pseudo code. Instead of using an existing path tracer and modifying it, we decided to write our own. Why? It is general, simple and self contained with no external dependencies and no complicated data structures and optimizations. At the highest level it works like this.

```
func do_path_tracing(do_gradient)
    if do_gradient then
        cost = 0
        clear_gradients()
    end if
    for each image pixel do
        pixel_radiance = 0
        for N samples do
            create initial ray R
            make_path(R)
            radiance = forward_pass() / N
            pixel_radiance += radiance
            if do_gradient then
                cost += cost(radiance)
                adjoint = dcost_drad(radiance)
                backward_path(adjoint)
            end if
        end if
    end for
end
```

When do_gradient==false we just perform a standard path trace. This is how it works. First, we create a path by intersecting the ray with the scene and spawning reflection vectors by sampling the corresponding PDFs. Then we traverse the path from the emitters to the receivers in a forward pass accumulating the radiance to generate a final radiance. To compute the gradients, we first compute the adjoint which is the Jacobian of the cost with respect to the radiance. Then, we perform a backward pass updating the adjoint and updating the gradient of each active control. The first step builds a path recursively.

```
func make_path(R)
    hit = int_scene(R)
    if hit!=0 and R->depth < MAX_DEPTH then
        u = unif(0,1)
        if u >= hit->absorb then
            dir = hit->sample(hit->normal,R->dir)
            R_new = make_ray(R->depth+1, hit->point, dir)
            make_path(R_new)
        end if
        path[R->depth]->{hit, dir_i, dir_o} = {hit, R->dir, dir}
    end if
end
```

Once the path is created, we perform a forward pass.

```
func forward_pass()
    N = path length
    hit = path[N-1]->hit
    radiance = hit->emission / hit->absorb
    for k=N-2 to k>=0 do
        {hit, dir_i, dir_o} = path[k]->{hit, dir_i, dir_o}
        path[k]->radiance = radiance
        radiance = hit->BSDF_D_PDF(hit->normal, dir_o, -dir_i) * radiance
    end for
    return radiance
end
```

After computing the cost and the adjoint we then perform a backward pass:

```
func backward_pass(adjoint)
    N = path length
    for k=0 to N-2 do
        {hit, dir_i, dir_o, radiance} = path[k]->{hit, dir_i, dir_o, radiance}
        radiance /= 1 - hit->absorb
        hit->update_BSDF_D_PDF_gradients(hit->normal, -dir_i, dir_o, radiance, adjoint)
        adjoint = hit->BSDF_D_PDF(hit->normal, -dir_i, dir_o) * adjoint
    end for
    hit = path[N-1]->hit
    radiance = hit->emission / hit->absorb
    hit->emission_gradient += radiance * adjoint
    adjoint = 0
end
```

The routines in italicized bold face must be provided for a particular BSDF/PDF. Here are the implementations for Lambert

```
func sample(normal, dir)
    u1, u2 = unif(0,1)
    c = sqrt(1-u1)
    s = sqrt(u1)
    (x,y,z) = (cos(2*PI*u2)*c, sin(2*PI*u2)*c, s)
    make_frame(normal, X, Y, Z)
    return x*X + y*Y + z*Z
end

func BSDF_D_PDF(normal, dir_i, dir_o)
    return diff_col
end

func update_BSDF_D_PDF_gradients(normal, dir_i, dir_o, radiance, adjoint)
    diff_gradient += radiance*adjoint
end
```

And for Phong-Blinn

```
func sample(normal, dir)
    u1, u2 = unif(0,1)
    t = pow(u1,2/(exponent+2))
    c = sqrt(1-t)
    s = sqrt(t)
    (x, y, z) = (cos(2*PI*u2)*c, sin(2*PI*u2)*c, s)
    make_frame(normal, X, Y, Z)
    N = x*X + y*Y + z*Z
    if dot(N,dir) < 0 then N = 2*dot(N,normal)*normal - N end if
    return 2*dot(N,dir)*N - dir
end
```

```
func BSDF_D_PDF(normal, dir_i, dir_o)
    dir_m = normalize(dir_i + dir_m)
    cos_m = dir_m*normal
    cos2_m = cos_m*cos_m
    sin_m = sqrt(1-dir_m*dir_m)
    return spec_col * sin_m
end

func update_BSDF_D_PDF_gradients(normal, dir_i, dir_o, radiance, adjoint)
    dir_m = normalize(dir_i + dir_m)
    cos_m = dir_m*normal
    cos2_m = cos_m*cos_m
    sin_m = sqrt(1-cos2_m)
    dsin_m = cos2_m * log(cos_m) / sin_m / (alpha+2)

    spec_gradient += sin_m * (radiance*adjoint)
    alpha_gradient += spec_col * dsin_m * (radiance*adjoint)
end
```

The function `make_frame()` generates an orthonormal frame from the normal:

```
func make_frame(N, X, Y, Z)
    Z = N
    X = Z + (0.1, 0.2, 0.3)
    Y = normalize(cross(Z,X))
    X = normalize(cross(Y,Z))
end
```

The orientation of the frame along the normal does not matter because we only consider isotropic BRDFs in this paper.